\newcommand{\refe}[1]{(\ref{#1})}

\newcommand{\refE}[1]{Eq.~(\ref{#1})}
\newcommand{\qav}[1]{\langle #1 \rangle }
\newcommand{\rem}[1]{}

\newcommand{\qv}{{\bf q}}
\newcommand{\kv}{{\bf k}}
\newcommand{\rv}{{\bf r}}
\newcommand{\pv}{{\bf p}}

\newcommand{\beq}{\begin{equation}}
\newcommand{\eeq}{\end{equation}}

\newcommand{\beqa}{\begin{eqnarray}}
\newcommand{\eeqa}{\end{eqnarray}}

\documentclass[showpacs,twocolumn,prb]{revtex4}

\usepackage{psfig}
\usepackage{graphicx}

\begin{document}

\title{Subgap noise of a superconductor-normal-metal tunnel interface}

\author{F. Pistolesi, G. Bignon, and F.~W.~J. Hekking}

\affiliation{
Laboratoire de Physique et Mod\'elisation des Milieux Condens\'es\\
Centre Nationale de la Recherche Scientifique and
Universit\'e Joseph Fourier\\
B.P. 166, 38042 Grenoble, France
}

\date{\today}

\begin{abstract}
It is well established that the subgap conductivity through a
normal-metal-insulator-superconductor (NIS) tunnel junction  is
strongly affected by interference of electron waves scattered by
impurities.
In this paper we investigate how the same phenomenon affects the low
frequency current noise,  $S$, for voltages $V$ and temperatures $T$
much smaller than the superconducting gap.
If the normal metal is at equilibrium we find that the simple relation
$S(V,T) = 4e\,{\rm coth} (eV/T) I(V,T) $ holds quite generally even for
non-linear $I$-$V$ characteristics.
Only when the normal metal is out of equilibrium, noise and current become independent.
Their ratio, the Fano factor, depends then on the details of the layout.
\end{abstract}

\pacs{ 74.40.+k, 74.45.+c, 74.50.+r
%
}

\maketitle

\section{Introduction}

Recently, a great progress has been achieved in the theoretical
understanding of current fluctuations in mesoscopic normal-metal
superconductor (N-S) hybrid systems.\cite{Buttiker,NazarovBook,RecentPRLS}
This progress has been boosted by the development of simple techniques
to calculate the full counting statistics of quantum charge
transfer.\cite{Levitov00,NazarovFCS00,MuzKhm}
In particular, the current fluctuations in a diffusive wire in good
contact with a superconductor have been calculated taking into account
the proximity effect for any voltage and temperature below the gap.
\cite{BelzigNazarovPRL2001}
With proximity effect we mean here the presence of a space
dependent coherent propagation in the normal metal of electrons
originating from the superconductor.
The opposite limit of a tunnel junction between a diffusive metal and
a superconductor has been less investigated, partially because only
very recently it has become possible to measure current noise in tunnel
junctions.\cite{CEArecent}
Theoretically, current noise in a NIS junction has been considered
by Khlus  long ago,\cite{Khlus} but neglecting proximity effect.
Later, de Jong and Beenakker included the proximity effect at
vanishing temperature and voltage.\cite{JongBennaker}
The effect of a finite voltage was studied very recently using a
numerical approach.\cite{numeric}
More complicated structures with several tunnel barriers have also
been considered. In some limits these can be reduced to a single
dominating NIS junction with a complex normal
region.\cite{BeamSplitter,Samuelssonn}
Actually, one may expect that the noise at finite voltage and
temperature (of the order of the Thouless energy $E_{th}$)
depends on the spatial layout when a tunneling barrier is
present,\cite{NazarovPRL94} but this has not been investigated so far on
general grounds.

As a matter of fact, it has been shown by one of the authors and
Nazarov in Ref. \onlinecite{HN94}, that the subgap Andreev tunnel current is
strongly affected by the coherent scattering of electrons by
impurities near the junction region.
Two electrons originating from the superconductor with a difference in
energy of $\varepsilon$ can propagate on a length scale of the order
of $\xi_{\varepsilon}=\sqrt{D/\varepsilon}$ before dephasing ($D$ being
the diffusion coefficient and we set $\hbar$ and $k_B=1$ throughout
the paper).
If the relevant energy scale of the problem, {\em i.e.} the
voltage bias $V$ multiplied by the electron charge $e$ and the
temperature $T$ are sufficiently small, the resulting coherence
length $\xi_{corr} = \sqrt{D/{\rm max}\{eV,T\}}$ is much larger
than the mean free path $l$.
Thus at low temperatures and voltages the electron pairs are able
to ``see'' the spatial layout on a length scale given by $\xi_{corr}\gg l$.
Since electron pairs attempt many times to jump into the
superconductor before leaving the junction region, interference
enormously increases the current at low voltage bias and the
resulting conductivity depends strongly on the explicit
layout.\cite{HN94}
In this paper we investigate how the same coherent diffusion of
pairs of electrons determines the subgap current noise of a NIS tunnel
junction.

We find that a generalized Schottky relation holds for voltages and
temperatures smaller than the superconducting gap:
 \beq
   S(V,T) = 4e \, {\rm coth}(eV/T)  I(V,T)
   \/.
  \label{maineq}
\eeq
Due to the interplay between the proximity effect and the presence
of the barrier the current voltage characteristics is both
non-universal and non-linear.
However, according to \refE{maineq}, the {\em ratio}
$F=S/(2eI)= 2 {\rm coth}(eV/T)$,
known as Fano factor, is universal as long as the electron
distribution on the normal metal is at equilibrium.
In particular, for $T\ll eV$ the Fano factor is 2, indicating that
the elementary charge transfer is achieved by Cooper
pairs.

If the normal metal is driven out of equilibrium,
within our formalism $F$ can be calculated once the geometry is known.
One finds then that the noise and the current remain both
independent of each other {\em and} strongly dependent on the geometry.
We discuss a realistic example (cf. Ref. \onlinecite{fil})
where we predict  strong deviations from \refE{maineq}.

The paper is organized as follows.
In Section \ref{secTH} we set up the main equations and obtain
the result given in \refE{maineq}.
In Section \ref{secOE} we discuss the case when the normal metallic
reservoirs are out of equilibrium.
In this case \refE{maineq} does not hold and to obtain the
Fano factor the diffusive nature of transport will play an
important role.
%
%
Section \ref{conclusion} gives our conclusions.

\section{Tunnelling Hamiltonian approach}
\label{secTH}

Let us consider a normal metallic reservoir connected through a tunnel
junction to a superconductor.
This problem can be treated conveniently by perturbation theory
in the tunnelling amplitude with the standard model Hamiltonian
$ H=H_S+H_N+H_T $.
Here $H_N$ and $H_S$ describe the disordered
normal metal and superconductor:
\beqa
   H_N
    &=&
    \sum_{\kv \sigma} \xi_\kv
    c^\dag_{\kv\sigma} c^{\vphantom \dag}_{\kv \sigma} \ ,
    \\
  H_S
    &=&
    \sum_{\qv \sigma} \zeta_\kv
    d^{\dag}_{\qv \sigma} d^{\vphantom \dag}_{\qv\sigma}
    +
    \sum_{\qv} [\Delta
    d^\dag_{\qv\downarrow} d^\dag _{-\qv\uparrow}
    +
    \Delta^*
    d^{\vphantom\dag}_{-\qv\uparrow} d^{\vphantom \dag}_{\qv\downarrow}
    ] \ ,
\eeqa
 $\Delta$ is the superconducting order paramenter,
$c_{\kv \sigma}$ and $d_{\qv \sigma}$ are destruction
operators for the electrons on the normal and superconducting
side, respectively.
With $\kv$ and $\qv$ we indicate the eigenstates of the disordered
Hamiltonian with eigenvalues $\xi_\kv$ and $\zeta_\qv$,
$\qv$ and $-\qv$ indicate time reversed states.
The index $\sigma$ stands for the spin projection eigenvalue.
The tunnelling part of the Hamiltonian reads:
\beq
    H_{T} = \sum_{\kv,\qv, \sigma}
        [t_{\kv,\qv} c^\dag_{\kv\sigma} d^{\vphantom \dag}_{\qv\sigma} +
        t^*_{\kv,\qv} d^\dag_{\qv\sigma} c^{\vphantom \dag}_{\kv\sigma}]
        \, ,
        \label{Htunn}
 \eeq
where $t_{\kv \qv} = \int d \rv d\rv' t(\rv,\rv') \psi_\kv(\rv)
\psi_\qv^*(\rv')$, and $t(\rv,\rv')$ is the quantum
amplitude for tunnelling from point $\rv$ of the normal side to point
$\rv'$ of the superconducting one.
All the information on the disorder is in the eigenfunctions
$\psi_{\kv/\qv}$.
The difference of potential $V$ between the two electrodes can be
taken into account by the standard gauge transformation $\xi_\kv
\rightarrow \xi_\kv-eV$.

We are concerned only with sub-gap properties:
$eV,T \ll \Delta$.
No single particle states are available in the superconductor for
times longer than $1/\Delta$.
Thus the low frequency noise ($ \omega \ll \Delta$) will be determined
only by tunneling of pairs.
Single particle thermal excitations in the superconductor are
exponentially suppressed and are thus neglected.
To calculate the quantum amplitude for transferring one Cooper
pair to the normal metal we proceed as in Ref.~\onlinecite{HN94}.
This gives:
\beq
    A_{\kv\kv'} = \sum_{\pv} t^*_{\kv\pv} t^*_{\kv' -\pv}
        u_\pv v_\pv
        \left\{
        {1\over \xi_\kv-E_\pv} + {1\over \xi_{\kv'}-E_\pv}
        \right\}
    \,,
\eeq
where $\kv$ and $ \kv'$ indicate the final electron states
(assumed empty), $v^2_\pv=1-u^2_{\pv}=(1-\zeta_\pv/E_\pv)/2$ are
the BCS coherence factors, and $E_\pv=(\zeta_\pv^2+\Delta^2)^{1/2}$
is the superconducting spectrum.

The function $A$ gives the amplitude for the only possible elementary
process at low energies.
It is thus convenient to write an effective tunnelling Hamiltonian
in terms of these processes,
\beq
    H^{eff} = H_N+J+J^\dag
\quad {\rm with}\quad
    J = \sum_{\kv \kv'}
         A_{\kv\kv'} c^\dag_{\kv\uparrow} c^\dag_{\kv'\downarrow} b_o
        \label{Jeff}\, ,
\eeq
where $b_o=(2/N_S) \sum_\qv \phi_\qv
d_{\qv\uparrow}d_{-\qv\downarrow}$ is the destruction operator for
a Cooper pair in the condensate, $\phi_\qv=v_\qv/u_\qv$ is the
pair wavefunction and $N_S$ is the number of fermions on the
superconducting side.
Note that $b_o$ is not a true bosonic operator, since Cooper pairs
overlap.\cite{PistolesiStrinati}
In practice, since the superconductor is in the coherent phase,
the averages involving $b_o$ are trivial to perform (we neglect
collective-modes excitations, which are at higher
energy): $\qav{b_o} = \qav{b^\dag_o}= \qav{b_o^\dag b_o}= 1 $.
A similar Hamiltonian, but with a constant amplitude, has been
considered by many authors to study the influence of collective
modes on Andreev reflection from a phenomenological point
of view.\cite{Wen,Falci}
Here we found its form from the BCS microscopic model, the
resulting $H^{eff}$ is thus equivalent to the starting one
\refe{Htunn} at low energies.

The current operator\footnote{With the sign convention that current
is positive when flowing from the normal metal to the
superconductor.} can be obtained as usual by the time
derivative of $N=\sum_{\qv\sigma}  c^\dag_{\kv\sigma}
c^{\vphantom\dag }_{\kv \sigma}$:
\beq
   I(t)
   =  e \, \frac{dN}{dt} =  e i [H^{eff}_T,N]
   = - 2 e i [J(t)-J^\dag(t)] \ .
\eeq
We have now all the elements to calculate by perturbative
expansion in the new coupling $A$ ($\sim t^2$) the quantum average
of the current, $\qav{I}$, and of the current-noise-spectral
density:
\beq
    S(\omega=0) =
    \int_{-\infty}^{+\infty} dt
    [ \qav{ [I(t), I(0)]_+}-2\qav{I(t)}\qav{I(0)}]
    \,.
\eeq
Writing the evolution operator in the interaction representation
$U(t)=T\exp\{-i \int_{-\infty}^t dt' [J_o(t')+J_o^\dag(t')]\}$,
where $J_o(t)$ evolves according to $H_N$, one can readily
expand in $A$ and obtain in lowest order the following expression
for $I$ and $S$:
\beqa
    I(t=0)
    &=& 2e [N_\leftarrow- N_\rightarrow]
    \label{expI}
    \\
    S(\omega=0)
    &=& 8e^2 [N_\leftarrow+ N_\rightarrow]
    \label{expS}
\eeqa
where
\[
    N_\rightarrow =\int_{-\infty}^{\infty} \!\!\!\!\!\!\!\! dt'
    \qav{J_o(0)J_o^\dag(t')}
    \quad,\quad
    N_\leftarrow  = \int_{-\infty}^{\infty} \!\!\!\!\!\!\!\!dt'
    \qav{J_o^\dag(t')J_o(0)}
    \,.
\]
This result depends only on the fact that the current can be
described as a first order process in the effective tunnelling amplitude
$A$.
For the tunnelling case Levitov and Reznikov\cite{Levitovbipoissonian}
showed that the full counting statistics is bi-poissonian:
\beq
    \ln \chi(\lambda)
     =
     (e^{ 2ie\lambda} -1) N_\leftarrow +
     (e^{-2ie\lambda} -1) N_\rightarrow
     \label{FCS}
     \,,
\eeq
where $\chi$ is the generating function.
Taking successive derivatives with respect to $i\lambda$ of $\ln \chi$
and setting $\lambda=0$ afterwards one generates
current momenta of all orders.
It is readily verified that
$I= \partial \ln\chi/\partial (i \lambda)$, and
$S= 2 \partial^2 \ln\chi/\partial (i\lambda)^2$
coincide with \refe{expI}-\refe{expS}.
\refE{FCS} implies that $I$ and $S$ determine the full counting
statistics of the problem.
The physical picture is simple: the distribution of transmitted
Cooper pairs is given by a superposition of two Poissonians for
tunnelling from left to right and from right to left.

If the metallic side is at the thermodynamic equilibrium
the full dependence of noise on voltage and
temperature is given on general grounds by \refe{maineq}.
This relation can be proven by evaluating \refe{expI} and \refe{expS}
using the basis of eigenstates $|n\rangle$ of $H_N$ with
eigenvalues $E_n$. One can actually write $I=2e C_-$ and
$S=8e^2 C_+$ with
\beqa
    C_\pm &=&\sum_{n\,m}
    \int_{-\infty}^{+\infty}\!\!\!\!\!\!\!\!\!dt\,
     e^{2ieVt } {e^{-E_n/T} \over {\cal Z}}
    \times \nonumber \\
    &&\left(|J_{mn}|^2 e^{i\omega_{nm}t} \pm  |J_{nm}|^2  e^{-i\omega_{nm} t}\right)
       \label{Cpm}
\eeqa
where $\omega_{nm}=E_n-E_m$, ${\cal Z}=\sum_n e^{-E_n/T}$ is the partition function
and $J_{nm}=\langle n | J | m \rangle$.
By simple manipulations of \refE{Cpm} one obtains
\beq
   C_\pm =
    2 \pi \left(1 \pm e^{2 e V/T} \right)
   \sum_{n\,m} {e^{-E_n/T} \over {\cal Z}}
    |J_{nm}|^2 \delta(2eV-\omega_{nm})
    \,.
    \label{Cpm2}
\eeq
From \refE{Cpm2} the relation \refe{maineq} follows
directly.
Note that the proof is valid for any system at
thermal equilibrium, interactions in the normal
metal do not spoil the result.
For a generic tunnelling system the relation between
noise and current was shown by Sukhorukov and Loss\cite{SukoLoss}
and discussed recently by Levitov.\cite{LevitovErice}

One thus finds that non-linear dependence on the voltage
and temperature of the current will be found exactly
in the low frequency noise with a {\em universal} Fano factor of $F=S/(Ie) =2\coth(eV/T)$.
In the completely coherent case ($A$ independent of energy, {\em i.e.}
$eV \ll E_{th}$) the same equation has been proved in Ref. \onlinecite{Martin}.
In this paper we extend its validity to arbitrary ratio $eV/E_{th}$
provided $eV\ll\Delta$.
Few comments are in order.
\refE{maineq} can be regarded as a generalization of the fluctuation dissipation
theorem, since it holds for any value of the perturbing field $V$.\cite{SukoLoss}
We would point out also that in presence of interactions among
electrons the noise was calculated recently in Ref. \onlinecite{SLF} with
the limitation $eV \ll E_{th}$.
We mention that for a normal wire in good
contact with the superconductor the situation is much richer and
the full energy dependence of the Fano factor has been calculated\cite{BelzigNazarovPRL2001}
and measured.\cite{Jehl,Prober1,Prober2}
For this case an analytical expression that links the current to the noise
was proposed,\cite{Houzet} but it does not reproduce the
weak energy dependence of the differential Fano factor
[cf. \refE{defDiffFano} in the following].
Only very recently an analytical solution  of the generalized Usadel
equation introduced in Ref. \onlinecite{BelzigNazarovPRL2001}
has been obtained.\cite{HouzetPistolesi}

\begin{figure}
\centerline{\psfig{file=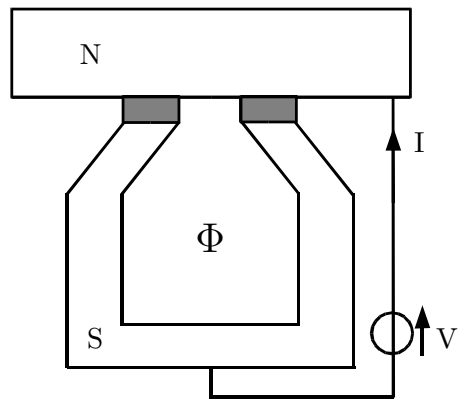,width=6cm}}
\caption{Experimental set up similar to Ref~\onlinecite{Gueron}.
The magnetic flux dependence of the current and noise can
be used to measure the doubling of the Fano factor with
high accuracy.
}
\label{FigExp}
\end{figure}

In order to emphasize the generality of \refE{maineq} let
us consider a non-trivial example.
In Ref.~\onlinecite{Gueron} the low-voltage anomalies due to the proximity
effect have been detected in a superconducting ring closed on a normal
metal through two tunnel junctions as shown in Fig. \refe{FigExp}.
Part of the subgap current through this structure is modulated by the
external magnetic field.
Using \refE{maineq} one can predict the following noise dependence on
the magnetic flux $\Phi$ through the ring: $S(\Phi)= 4e \,{\rm
coth}(eV/T) I(\Phi)$.
Since the $\Phi$ dependent part is given by the coherent propagation
of two electrons between the two junctions any single particle
contribution is eliminated from the outset.
This experiment could thus be used to test accurately the
validity of \refE{maineq} and in particular the doubling
of the charge at low temperature.

\section{Out of equilibrium reservoirs in diffusive junctions}
\label{secOE}

When the normal metal is not at equilibrium the simple
relation \refe{maineq} between $S$ and $I$ does not
hold any longer, and expressions \refe{expI} and \refe{expS} have  to be
evaluated explicitly.
We proceed by writing:
\beqa
  I &=&
     4\pi e \sum_{\kv \kv'} |A_{\kv \kv'}|^2
    \delta(\xi_\kv+\xi_{\kv'}-2eV) H_-(\xi_\kv,\xi_{\kv'})
    \,,
\nonumber
    \\
  S &=&
     16 \pi e^2 \sum_{\kv \kv'} |A_{\kv \kv'}|^2
    \delta(\xi_\kv+\xi_{\kv'}-2eV)
    H_+(\xi_\kv,\xi_{\kv'})
    \,,
    \nonumber
\eeqa
where:
\beq
   H_{\pm}(\xi,\xi') = [1-n(\xi)][1-n(\xi')] \pm  n(\xi)n(\xi')
   \,,
   \label{Hpm}
\eeq
and $n(\xi_\kv)=\qav{c^\dag_{\kv\sigma} c^{\vphantom \dag}_{\kv\sigma}}$.
It is convenient to introduce the relevant energy dependence by
defining:
\[
   A(\varepsilon,eV) =
   \sum_{\kv \kv'}
   \delta(\varepsilon/2+eV-\xi_\kv)
   \delta(-\varepsilon/2+eV-\xi_{\kv'})
   |A_{\kv \kv'}|^2 .
\]
We thus obtain:
\beq
    \left({ I \atop S}\right)
    =
     2 \pi e \!\!\!
    \int_{-\infty}^{\infty} \!\!\!\!\!\! d\varepsilon
    A(\varepsilon,eV)
    \left({
    H_-({\varepsilon\over 2}+eV,-{\varepsilon\over 2}+eV)
    \atop
    4 e H_+({\varepsilon\over 2}+eV,-{\varepsilon\over 2}+eV)  }
    \right)
    \,.
    \label{ISbetter}
\eeq
If the normal metal is in thermal equilibrium $n(\xi)$ is the
Fermi distribution $f_F(\xi)$.
In that case $H_+$ and $H_-$ are related
by the simple relation:
\beq
 {
 H_+(\varepsilon/2+eV,-\varepsilon/2+eV)
 \over
 H_-(\varepsilon/2+eV,-\varepsilon/2+eV)
 }
=
{\rm coth}\left({eV \over T}\right)
\ ,
\label{eqcoth}
\eeq
and we recover again \refE{maineq}.
If $n(\xi)\neq f_F(\xi)$, \refE{eqcoth} does not
hold and $S$ and $I$ become independent.
Indeed, the Fano factor ceases to be universal and gives new
information on the system.
The technique developed so far enables calculation of $F$
also in this case through \refE{ISbetter}, however now
we need to specify the explicit form of $A(\varepsilon)$.
We follow Refs.~\onlinecite{HN94} to evaluate it:
\beqa
  \lefteqn{A(\varepsilon, eV) =
    \int d\zeta d\zeta' d\xi d\xi'
    F(\zeta;\xi,\xi')F(\zeta';\xi,\xi')
    \times} \nonumber\\
    &&
    \Xi(\zeta,\zeta';\xi,\xi')
   \delta(\varepsilon/2+eV-\xi)
   \delta(\varepsilon/2-eV+\xi')
   \nonumber
\eeqa
 where
 $F(\zeta;\xi,\xi') = u(\zeta)v(\zeta)
 \{(\xi+eV-E(\zeta))^{-1}+(\xi'+eV-E(\zeta))^{-1}\}
$,
%
\beqa
    \lefteqn{\Xi[\zeta,\zeta',\xi,\xi'] =
    \!\int\!\! d^3\rv_1 \dots d^3 \rv_4
    \int \!\! d^3\rv'_1  \dots d^3 \rv'_4 \times }
    \nonumber \\
    &&
    \qav{K_{\xi}(\rv_1,\rv_3) K_{\xi'}(\rv_2,\rv_4)}_{imp}
    \qav{K_{\zeta}(\rv'_2,\rv'_1) K_{\zeta'}(\rv'_4,\rv'_3)}_{imp}
    \times
    \nonumber \\
    &&
    t^*(\rv_1,\rv'_1) t^*(\rv_2,\rv'_2)
    t(\rv_3,\rv'_3) t(\rv_4,\rv'_4)
    \,.
    \label{BigEq}
\eeqa
%
%
Merely $ K_\xi(\rv_1,\rv_2) = \sum_\kv \delta(\xi-\xi_\kv)
\psi_\kv(\rv_1)\psi^*_\kv(\rv_2)$ is the single particle
propagator and $\qav{\dots}_{imp}$ indicates impurity average.
Since the tunnelling is most probable only near the surface of the
junctions, we can set $\rv_i=\rv'_i$ and restrict them
to lie on the surface.
The impurity averages can be performed within the usual approximation
$k_F l \gg 1$ ($k_F$ Fermi momentum).
The main contribution to \refE{BigEq} is given by the ``Cooperon''
diagrams, associated to the coherent propagation of two interfering
electrons with small energy difference $\varepsilon$ and small total
momentum.
The range of the Cooperon propagator is $\xi_\varepsilon$, and thus it
is much larger than the range of $K_\xi(\rv,\rv')$ which is $l$.
For this reason among the integrals over $d^2 \rv$ two dominant
terms can be singled out:
(a) the term where $\rv_1\approx \rv_2$ and $\rv_3 \approx \rv_4$
with interference occurring in the normal metal
and
(b) the term where $\rv_1\approx \rv_3$ and $\rv_2 \approx \rv_4$
with interference occurring in the superconductor.
Since in Ref.~\onlinecite{HN94} it has been shown that the second term is
always smaller than the first by a factor $eV/\Delta$
we consider only case (a).
The dominant contribution to $\Xi$ can thus be written
as follows:
$\Xi[\zeta,\zeta';\xi,\xi']\approx\Xi(\xi-\xi')$ and
\beq
   \Xi(\varepsilon) = {G_T^2\over 32 \pi^3 e^4 \nu_N {\cal S}^2}
   \int d^2\rv_1 d^2 \rv_2
   [P_{\varepsilon}(\rv_1,\rv_2)+P_{-\varepsilon}(\rv_1,\rv_2)]
   \,.
   \label{CsiEq}
\eeq
Here $G_T$ is the conductance in the normal phase, ${\cal S}$ is
the surface area of the junction, $\nu_N$ is density of states of
the normal metal (per spin) and $P_\varepsilon(\rv)$ satisfies the
diffusion equation in the normal metal:
\beq
   (-D \nabla_1^2 -
   i\varepsilon) P_\varepsilon(\rv_1,\rv_2)
   = \delta^3(\rv_1-\rv_2)
   \,.
   \label{diffusionEq}
\eeq
The integration over $\zeta$ and $\zeta'$ for $\xi,\xi', eV \ll \Delta $
can be done and gives $\pi^2$, we are thus left with an explicit
expression for $A(\varepsilon)=\pi^2 \Xi(\varepsilon)$ which can
be calculated if the geometry of the system is known.
Given $A$ and $n(\xi)$ the noise and the current can be found
explicitly from \refe{ISbetter} and \refe{Hpm}.

\subsection{An explicit example: a wire out of equilibrium}

We are now in a position to calculate both current and noise in a
non-equilibrium system.
Let us consider a realistic example: In Ref.~\onlinecite{fil} the
non-equilibrium electron distribution for a small metallic wire of
length $L$ has
been measured by using a tunnel junction between the wire (at
different positions $x$ along the wire) and a supercondutor
(see also inset of Fig. \ref{Circuit} in the following).
In that case the quasiparticle current was measured, but the subgap
current and noise in the same configuration could also be measured.
When $L$ is shorter than the inelastic mean free path the electron distribution
along the wire is given by the diffusion equation:\cite{fil}
\beq
   n(x,\xi) = f_F(\xi+eU)[1-x/L] + f_F(\xi) x/L
   \label{anomalousn}
\eeq
where $U$ is the difference of potential between the two
normal reservoirs.
This leads to a double discontinuity of the distribution function at
$T=0$.

Since the transverse dimension $w$ of the wire is much smaller than
$\xi_{corr}$, we can use the one dimensional
diffusion equation.
Following Ref. \onlinecite{Altshuler} we impose the boundary condition on the
propagator $P$ corresponding to a good contact with a reservoir.
This means that $P_\varepsilon(x_1, x_2)$ should vanish when $x_1=0$ or $L$.
In the dimensionless variables $u=x/L$, \refE{diffusionEq} becomes
\beq
    {d^2P\over du^2_1}(u_1,u_2) + i {\varepsilon \over E_{th}^W}  P(u_1,u_2)
    =
    -{ L \over w^2 D} \delta(u_1-u_2)
\eeq
with the boundary conditions $P(0,u_2)=P(1,u_2)=0$ and
where $E_{th}^W = D/L^2$ is the Thouless energy of the wire.
The solution can be written as a sum of a special
solution of the complete equation plus a linear combination of the two
linearly independent solutions of the homogeneous equation.
The coefficients of this combination can be chosen in such a way as to
fulfill boundary conditions.
The solution reads:
\beqa
\lefteqn{P(u_1,u_2) = {L\over 2 D w^2 z}
\left[
e^{-z|u_1-u_2|} \phantom{ a^a \over b^a}
\right.} \nonumber
\\
&&
\left. -{
e^{z(u_1+u_2)}+e^{z(u_1+u_2-2)}-e^{z(u_1-u_2)}-e^{-z(u_1-u_2)}
\over
e^{2z}-1 }
\right] \,, \nonumber \\
\label{soluP}
\eeqa
where, for $\varepsilon>0$ , $z=\sqrt{\varepsilon/E_{th}^W} e^{-i \pi/4}$.
Inserting \refE{soluP} into \refE{CsiEq} and assuming that the size
of the junction is small, we have
\beq
    \Xi(\varepsilon) = {G_T^2  L \over 16 \pi^3 e^4 \nu_N D w^2}\, \phi(\varepsilon)\,,
\eeq
where
\beq
     \phi(\varepsilon)=
        {\rm Re}
        \left[
        {
        \sinh(z u_0) \sinh(z(1-u_0))
        \over
        z \sinh(z)
        }
        \right]
        \,,
\eeq
and $u_0$ indicates the position of the junction along the wire.
Introducing the conductance of the wire
$G_W= 2  w^2e^2 D \nu_N/L$ and using \refE{ISbetter}, the  expression for
the current simplifies to
\beq
    I = {G_T^2 \over 4 e G_W }
        \int \!\!  d\varepsilon \, \phi(\varepsilon) H_-(\varepsilon/2-eV,-\varepsilon/2-eV)
        \,.
        \label{IFilo}
\eeq
The noise obeys  an identical expression with $H_- \rightarrow 4 e H_+$.
Note that in the limit of an infinite wire the function
$\phi(\varepsilon) \sim 1/\sqrt{\varepsilon}$ diverges at low
energy, as expected for one dimensional diffusion.
Substituting \refE{Hpm} with \refE{anomalousn} into
\refE{IFilo} one obtains current and noise for any
temperature, voltage, and position along the wire.
Let us consider for simplicity one specific case:
zero temperature and voltage  $0 \leq V \leq U/2$.
It is not difficult to show that in this case
\beq
    I = {G_T^2 \over 2 e G_W} \phi_-\,,  \quad {\rm and } \quad
    S = 2 { G_T^2 \over  G_W} \phi_+
    \label{ISprediction}
\eeq
with
\beq
    \phi_\pm
    =
    (1-u_0) \int_{2eV}^{2e(U-V)} \!\!\!\!\!\!\!\!\!\!\!\!\!\!\!\!\!\!
      \phi( \varepsilon)  d \varepsilon +
    [ (1-u_0)^2 \pm u_0^2  ] \int_{0}^{2eV}\!\!\!\!\!\! \!\!\
     \phi( \varepsilon)  d\varepsilon
     \,.
\eeq
Differentiating with respect to $V$ Eqs. \refe{ISprediction} gives
the differential Fano factor
\beq
    {\cal F}(eV)  \equiv {(dS/dV) \over ( 2e\,  dI/dV)}\,.
    \label{defDiffFano}
\eeq
In our case ${\cal F}$ is given by the following expression
\beq
 {\cal F}(eV)  =
    2
    {  (1-u_0)\phi(2eU-2eV) - u_0\,(2 u_0-1)\phi(2eV)
    \over
    (1-u_0) \phi(2eU-2eV)+u_0 \phi(2eV)
    }
    \,.
    \label{diffFano}
 \eeq

\begin{figure}[tbh]
\centerline{
\psfig{file=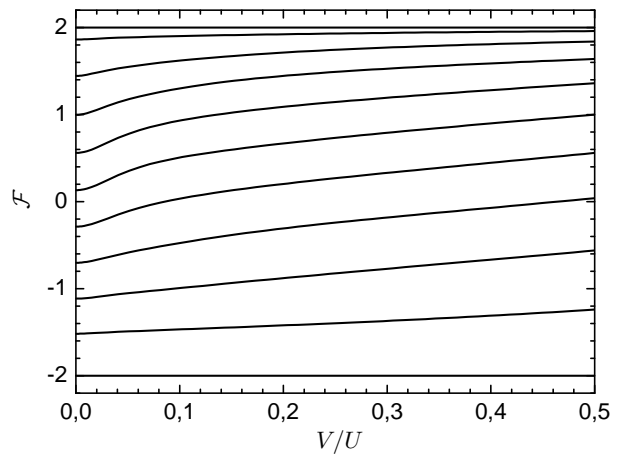,width=8.cm,angle=0}
}
\caption{
Differential Fano factor ${\cal F}$ as a function of $V/U$.
Different curves represent different values of $u_0$ going from 0 to 1 in step of 0.1
starting from the top.
Departure from the value 2 or -2 is due to the energy dependence
of the tunnelling matrix element.
}
\label{figFano}
\end{figure}

Note that normally the Fano factor is positive since
it corresponds to the (positive) current noise divided
by the absolute value of the current.
The differential Fano factor is then usually
defined in such a way that a factor sign$(I)$
is implicit in the definition.
This makes it usually positive, since current
and the noise are supposed to increase with
the voltage bias.
But in our case the wire is off-equilibrium,
thus the voltage $V$ has not the usual meaning
of bias voltage between two systems at local
equilibrium.
Along the wire it is not even possible to define
a chemical potential, since fermions are not
at equilibrium.
Thus the concept of difference of potential between the
superconductor and the wire at the position of the
contact is ill-defined.
The potential $V$ is nevertheless well defined
and can be used to study the evolution of the
conductance or of the differential noise.
Clearly the sign of the current needs not be
the same as that of V.
For this reason we use the definition
\refe{defDiffFano} as it is, without changing the sign
according to the direction of the current.
This allows a more simple representation in
Fig. \ref{figFano}.
One should keep in mind
that the change of sign has no special meaning in
this case, since increasing the voltage bias
$V$ can well decrease the current and the noise
at through the SIN junction.

Let us discuss briefly some simple limiting cases of
\refE{diffFano}.
For $u_0=0$ and $1$, when the junction is at
the extremities of the wire, \refE{diffFano} gives
${\cal F}=\pm 2$.
This is expected from \refE{maineq} since in this
case the normal metal is actually at equilibrium.
The sign change is simply due to fact that for
$x=0$ the potential of the wire $U$ is greater than the
potential of the superconducting point $V$, while for $x=L$
the situation is reversed.
For $0<u_0<1$, ${\cal F}$ is in general different
from 2.
An interesting point is the middle of the wire $u_0=1/2$.
In this case we have:
\beq
 {\cal F}(eV,u_0=1/2)  =
    2
    {\phi(2eU-2eV)
    \over
    \phi(2eV)+\phi(2eU-2eV)
    }
    \,.
\eeq
For $V=U/2$ we then obtain ${\cal F}=1$ exactly for any form
of $\phi(\varepsilon)$, while for $V=0$ ${\cal F}$ should become
very small on general grounds, since we have seen that for $\varepsilon\rightarrow 0$
the function $\phi(\varepsilon)$ is expected to diverge in the infinite system.

%
%

For arbitrary values of $u_0$ one can see a crossover
between these limiting behaviors.
We report in Fig. \ref{figFano}
the results predicted by \refE{diffFano} for $U = 200 E_{th}^W$
which is a typical value for experiments.
We would like to emphasize that the whole energy
dependence seen in this plot is due to interference
of electronic waves, since it stems from
the $\varepsilon$-dependence of $\phi(\varepsilon)$.

\subsection{Comparison with circuit theory approach}

\newcommand{\Gu}{\check{\cal G}}
\newcommand{\Ac}{\check{\cal A}}

It can be useful to obtain current and
noise with an alternative technique, often called circuit
theory, developed by Nazarov and
coworkers. \cite{NazarovFCS00,RecentPRLS,BelzigNazarovPRL2001,Belzig}
The assumptions behind circuit theory are the same as the ones that
have been used to arrive to \refE{diffFano}, but its scope is much larger,
since it can be applied for arbitrary transparency of
the interface.
On the other hand its implementation is often numerical and
can become extremely cumbersome in higher dimension.
In the case at hand of a wire the problem can be solved rather
easily and it is worth to compare the results of the two techniques.

We follow Belzig and Nazarov\cite{BelzigNazarovPRL2001} in
modelling the wire within semiclassical Green's function technique.
The wire is describled by $N$ nodes, each node represents a
part of the wire small enough such that the inhomogeneities inside
can be neglected.
Each node is thus completely described by one Usadel 4x4 matrix $\Gu$
in the Nambu$(\,\hat{}\,)$-Keldysh$(\,\bar{}\,)$ space
with the condition $\Gu^2=1$.
The nodes are connected by tunnel junctions with conductance
$g_T = (N+1)G_W $ in such a way that the total conductance
of the wire is $G_W$.
Kuprianov-Lukichev\cite{KL}  boundary conditions plus the
decoherence term in Usadel equation give the following condition
 to be satisfied at each node $i$
\beq
    [\Ac_i,  \Gu_i]=0
    \label{eqG1}
\eeq
where
\beq
    \Ac_i = g_T\,(\Gu_{i+1}+\Gu_{i-1}) - 2 i {\varepsilon\over N\, E_{th}^W} \check \tau_3 ,
    \label{eqG}
\eeq
$\check \tau_3=\hat\tau_3 \otimes \bar 1$, and
$\bar\sigma_i,\hat\tau_{j(i,j=1,2,3)}$ are Pauli matrices
($\bar\sigma$ will appear in the following).
The boundary conditions at the two extremities are given
by the bulk solution for the normal metal at equilibrium:
$\Gu_{N}^{0} = \hat\tau_3\otimes\bar\sigma_3+
(f_{T0}+f_{L0}\hat\tau_3)\otimes(\bar\sigma_1+i\bar\sigma_2) $
with
$f_{L0}=1-f_+-f_-$, $f_{T0}=f_--f_+$,
$f_{\pm}(\varepsilon)=f(\varepsilon\pm eU_{L/R})$, $f$ is the Fermi
function at temperature $T$, and $U_{L/R}$ is the voltage bias
at (L) $x=0$ and at (R) $x=L$.
\refE{eqG1} has a simple analytical solution:
$\Gu=\Ac_i/\sqrt{\Ac_i^2}$, and the full set of equations can be
solved by iteration leading to an explicit numerical
value for $\Gu_i$ at each node.

\begin{figure}[tbh]
\centerline{
\psfig{file=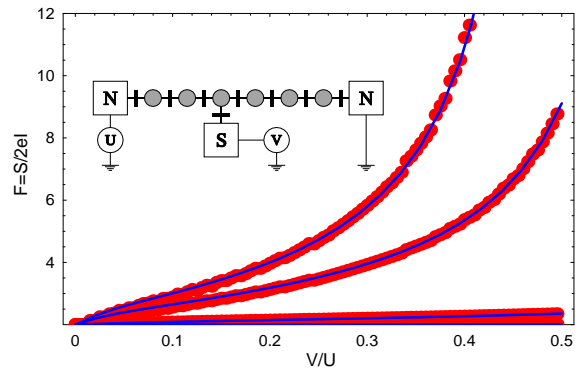,width=8.cm,angle=0}
}
\caption{
Fano factor as a function of $V/U$ for $x/L=9/18,$ 8/18, 4/18, and 1/18,
from top to bottom.
Full lines are obtained with \refE{ISprediction}.
The dots indicate the result obtained with circuit theory.
The inset shows the circuit used in the calculation.
}
\label{Circuit}
\end{figure}

This describes a normal wire in good contact with two
normal reservoirs kept at some voltage bias $U_L$ and $U_R$ with respect
to (an arbitrary) ground.
We add now the superconducting tunnel contact
with conductance $G_T$ at node $s$, with $1 \leq  s \leq N$.
Since it is easier technically to keep the superconductor
at zero voltage with respect to ground, we set $U_L=U-V$ and $U_R=-V$.
The tunnel contact with the superconductor modifies
the conditions \refe{eqG} for node $s$ in the following
way:
\beq
   \Ac_s
   =
   g_T\,(\Gu_{s+1}+\Gu_{s-1})+G_T\, \Gu_S(\chi)
   - 2 i {\varepsilon \over N\, E_{th}^W} \check \tau_3
   \,.
\eeq
(Since $G_T \ll g_T$ we can neglect corrections to the Thouless
energy due to the presence of the superconducting reservoir.)
The Green's function $\Gu_S(\chi)$ is the
superconducting bulk solution of the
semiclassical equations, modified by the
introduction of a counting field $\chi$.
This means that
$
    \Gu_S(\chi)
    =
    e^{-\frac{i}{2}\chi\check\tau_{K}}\,
    \Gu_{S}^{0} \,
    e^{\frac{i}{2}\chi\check\tau_{K}}
$
with
$\Gu_{S}^{0} =\hat\tau_2 \otimes \bar 1$
(since all energies are smaller than the superconducting
gap) and $\check\tau_{K}=\hat\tau_3\otimes\bar\sigma_1$.
The circuit is shown in the inset of Fig. \ref{Circuit}.
Once the solution to the system of equations has been found
it is possible to obtain the current, the noise, and,
in principle, all higher cumulants of the current flowing
through the SIN junction by evaluating
\beq
    J(\chi) = -{G_T  \over 8e } \int d \varepsilon \,
    {\rm Tr} \left\{ \check\tau_K [\Gu_s,\Gu_S] \right\}
    \,.
\eeq
The current is given by $J(\chi=0)$
while the noise is proportional to its first
derivative with respect to $\chi$:
$S=-2ie \partial J /\partial \chi$.
Numerically one can obtain both at the same time,
since for very small $\chi$ the first is proportional to the real
part of $J(\chi)$ while the second to its imaginary part.

We have thus found numerically the current and the noise
for $U=200 E_{th}^W$ and different values of $V$.
The numerical results presented refers to 18 nodes.
The tunnel conductance $G_T$ was chosen much smaller
than $g_T$ ($g_T/G_T \sim 10^3$) and we have verified that both
current and noise scale with $G_T^2$ as predicted by
\refE{ISprediction}.
%

%
%

In Fig. \ref{Circuit} we show the Fano factor as a function of $V/U$
($U=200 E_{th}^W$, $V<U/2$) for different value of $s$,
the node in contact with the superconductor.
The agreement between the two approaches is rather good.

A strong effect of the non equilibrium distribution
is the divergence of the Fano factor for certain values
of $V$ or $u_0$, while at equilibrium and at zero temperature
\refE{maineq} predicts invariably $F=2$.
In the case shown in Fig. \ref{Circuit} the divergence appears
for $x=L/2$ and $V=U/2$.
This divergence can be easily understood, since at these values
of the bias and position along the wire the current
vanishes with $N_\rightarrow = N_\leftarrow \neq 0$ [cf. \refE{expI}]
such that the resulting $S$ is non vanishing [cf. \refE{expS}].
Even if this is a large effect it is mainly due to the
fact that the system is out of equilibrium, and it is thus not
a direct signature of the importance of the interference
in the noise.
Interference appears instead directly in the
differential Fano factor $\cal F$.

\begin{figure}[tbh]
\centerline{
\psfig{file=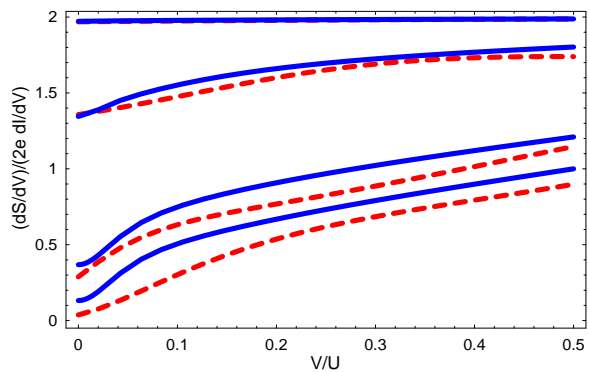,width=8.cm,angle=0}
}
\caption{
Comparison of the differential Fano factor $\cal F$ given
by \refE{diffFano} (full lines) and by the numerical circuit theory
simulation (dashed lines).
The four dashed curves are obtained for $U=200 E_{th}^W$ and $u_0=1/18$, 4/18, 8/18,
and 9/18 ($N=18$).}
\label{NumericdiffFano}
\end{figure}

A more stringent test on the validity of both
approaches is thus to compare $ {\cal F}(eV)$.
Our rather simplified numerical approach to the circuit theory
allows to extract the derivatives of $S$ and $I$ with
a limited accuracy.
Nevertheless the agreement of the analytical expression
\refe{diffFano} and the numerical results shown
in Fig. \ref{NumericdiffFano} is reasonable, and
it remains within the numerical error of the circuit theory
calculation.
%

There are actually other reasons for possible
discrepancies between the two approaches.
In the analytical calculation we have not taken into account that
the Fermi distribution changes along the wire due to the electric field,
while this effect is included in the circuit theory approach.

%
%

\section{Conclusions}
\label{conclusion}

In conclusion we have found a general expression that links the subgap
noise and the current in a NIS junction.
The interference of electron pairs leads to a non-linear
dependence of current and noise on voltage, but their ratio
is fixed by \refE{maineq} as long as
the electrons in the normal metal are in equilibrium.
Out of equilibrium, the Fano factor becomes non
universal and we have computed it in a feasible
experiment.
In particular we have studied the differential Fano factor, which is
a direct measure of the importance of interference of electronic
waves.
The validity of \refE{maineq} is quite general, it does not
depend on the interactions in the normal metal, for instance.
Thus detecting a departure from this prediction can be a strong
experimental indication that the normal metal is out of equilibrium.

From the technical point of view we have verified that the
tunnelling approach agrees with the circuit theory approach.
Both stem from the semiclassical theory of current fluctuations,
but it is clear that the tunnelling calculation is often much
simpler than solving the full Usadel equations.
It is thus useful to verify that it can give reliable
quantitative prediction for both the current and the noise
in a specific example.

Let us briefly discuss a possible experimental test of \refE{maineq}.
Due to the large barriers induced by oxides, experiments on shot noise
in tunnel junctions are not as developed as they are for the transparent
junctions.
The only data available at this moment\cite{CEArecent}
appear to agree reasonably well with our prediction.
Indeed, current and noise both show a strong non-linear behavior, but
their ratio follows the simple relation \refe{maineq}.

We are indebted to M. Houzet for useful discussions.
We acknowledge financial support
from CNRS/ATIP-JC 2002 and Institut Universitaire de France.
%

%


\begin{thebibliography}{17}






\bibitem{Buttiker} For a recent review see
Y.M. Blanter and M. B\"uttiker,
Phys. Rep. {\bf 336}, 1 (2000).

\bibitem{NazarovBook}
{\em Quantum noise in mesoscopic systems}, edited by Y. V. Nazarov,
Kluwer, Dordrecht (2003).


\bibitem{RecentPRLS}
W. Belzig and Y.V. Nazarov, Phys. Rev. Lett. {\bf 87}, 197006 (2001);
Y. V. Nazarov and D.A. Bagrets, ibid {\bf 88}, 196801 (2002).


\bibitem{Levitov00}
L.S. Levitov, H.W. Lee, and  G.B Levitov,
J. Math. Phys. {\bf 37}, 4845 (1996).

\bibitem{NazarovFCS00}
Y. V. Nazarov, Ann. Phys. (Leipzig) {\bf 8}, SI-193 (1999).

\bibitem{MuzKhm}
B.A. Muzykantskii and D.E. Khmelnitskii,
Phy. Rev. B {\bf 50}, 3982 (1994).


\bibitem{BelzigNazarovPRL2001}
W. Belzig and Y.V. Nazarov,
Phys. Rev. Lett. {\bf 87}, 067006 (2001).


\bibitem{CEArecent}
F. Lefloch, C. Hoffmann, M. Sanquer, and D. Quirion,
Phys. Rev. Lett. {\bf 90}, 067002 (2003).

\bibitem{Khlus}
V.A. Khlus, Sov. Phys. JETP {\bf 66}, 1243 (1987).


\bibitem{JongBennaker}
M.J.M. de  Jong and C.W.J. Beenakker,
Phys. Rev. B {\bf 49}, 16070 (1994).


\bibitem{numeric}

M.P.V. Stenberg and T.T. Heikkil\"a,
Phys. Rev. B, {\bf 66}, 144504 (2002).


\bibitem{BeamSplitter}
J. B\"orlin, W. Belzig, and C. Bruder,
Phys. Rev. Lett. {\bf 88} 197001 (2002);
%
P. Samuelsson and M. B\"uttiker,
ibid., {\bf 89}, 046601 (2002).

\bibitem{Samuelssonn}
P. Samuelsson,
Phys. Rev. B 67, 054508 (2003).


\bibitem{NazarovPRL94}
Y.V. Nazarov,
Phys. Rev. Lett. {\bf 73}, 134 (1994).


\bibitem{HN94}
F.W.J. Hekking and Y.V. Nazarov,
Phys. Rev. Lett. 71, 1625-1628 (1993)
and Phys. Rev. B {\bf 49} 6847 (1994).


\bibitem{fil}
H. Pothier, S. Gu\'eron, N.O. Birge, D. Esteve, M.H. Devoret,
Z. Phys. B {\bf 104}, 178, (1997).


\bibitem{PistolesiStrinati}
F. Pistolesi and G.C. Strinati,
Phys. Rev. B {\bf 53}, 15168 (1996).

\bibitem{Wen}
Y.B. Kim and X.-G. Wen,
Phys. Rev. B {\bf 48}, 6319 (1993).


\bibitem{Falci}
G. Falci, R. Fazio, A. Tagliacozzo, G. Giaquinta,
Eur. Phys. Lett. {\bf 30}, 169 (1995).


\bibitem{Levitovbipoissonian}
L.S. Levitov and M. Reznikov,
{\tt cond-mat/0111057} (unpublished).


\bibitem{SukoLoss}
E.V. Sukhorukov and D. Loss, in {em Electronic Correlations: From Meso- to Nano-Physics},
edited by G. Montambaux and T. Martin, XXXVI Rencontre de Moriond; also
{\tt cond-mat/0106307}.

\bibitem{LevitovErice}
L.S. Levitov, in {\em New directions in Mesoscopic Physics (Towards Nanoscience)},
edited by R. Fazio, V.F. Gantmakher, and Y. Imry, Kluwer, Dordrecht (2003).
Also {\tt cond-mat/0210284}.


\bibitem{Martin}
Th. Martin, Physics Lett. A {\bf 220}, 137 (1996).

\bibitem{SLF}
M.A. Skvortsov, A.I. Larkin, and M.V. Feigel'man
Phys. Rev. B, {\bf 63}, 134507 (2001).


\bibitem{Jehl}  X.\ Jehl, M.\ Sanquer, R.\ Calemczuk, and D.\ Mailly,
Nature \textbf{405}, 50 (2000).

\bibitem{Prober1}
A.~A.\ Kozhevnikov, R.~J.\ Schoelkopf, and D.~E.
Prober, Phys. Rev. Lett. \textbf{84}, 3398 (2000).

\bibitem{Prober2}
B. Reulet, A.A. Kozhevnikov,
D.E. Prober, W. Belzig, Yu.V. Nazarov,
Phys. Rev. Lett. {\bf 90}, 066601 (2003).


\bibitem{Houzet}
M. Houzet and V.P. Mineev,
Phys. Rev. B {\bf 67}, 184524 (2003).

\bibitem{HouzetPistolesi}
M. Houzet and F. Pistolesi, Phys. Rev. Lett. {\bf 92}, 107004 (2004).


\bibitem{Gueron}
H. Pothier, S. Gu\'eron, D. Esteve, M.H. Devoret,
Phys. Rev. Lett.  {\bf 73}, 2488, (1993).


\bibitem{Altshuler}
B.~L.\ Altshuler and A.~G.\ Aronov, in {\em Electron-electron
interactions in disordered systems}, Eds. A.~L.\ Efros and M.~
Pollak, (North-Holland, Amsterdam) (1985).

\bibitem{Belzig} W. Belzig, in {\em Quantum Noise},
edited by Yu. V. Nazarov and Ya. M. Blanter (Kluwer)
also cond-mat/0210125, and reference
therein.

\bibitem{KL}
M.~Yu.\ Kuprianov and V.~F.\ Lukichev, Sov. Phys. JETP
\textbf{67}, 1163 (1988).



\end{thebibliography}

\end{document}